\documentclass[11pt]{article}
\usepackage{graphicx,amssymb,amsmath,amsthm}

\usepackage{algorithm}
\usepackage{algorithmic}

\usepackage{latexsym}
\usepackage{epic}
\usepackage{eepic}

\newtheorem{theorem}{Theorem}[section]
\newtheorem{lemma}[theorem]{Lemma}

\newtheorem{proposition}[theorem]{Proposition}
\newtheorem{corollary}[theorem]{Corollary}

\theoremstyle{definition}
\newtheorem{definition}[theorem]{Definition}
\newtheorem{definitions}[theorem]{Definitions}
\newtheorem{example}[theorem]{Example}

\newcommand{\ignore}[1]{}

\begin{document}

\title{Spanning connectivity games}

\author{
\em Haris Aziz
\and \em Oded Lashich
\and \em Mike Paterson
\and \em Rahul Savani
}

\date{} 

\maketitle

\begin{abstract}
The Banzhaf index, Shapley-Shubik index and other voting power
indices measure the importance of a player in a coalitional
game. We consider a simple coalitional game called the
spanning connectivity game (SCG) based on an undirected,
unweighted
multigraph, where edges are players. 
We examine the computational complexity of computing the 
voting power indices of edges in the SCG. 
It is shown that computing Banzhaf values and Shapley-Shubik indices is \#P-complete for SCGs. 
Interestingly, Holler indices and Deegan-Packel indices 
can be computed in polynomial time.
Among other results, it is proved that Banzhaf indices 
can be computed in polynomial time for graphs with bounded treewidth. It is also shown that for any reasonable representation of a simple game, a polynomial time algorithm to compute the Shapley-Shubik indices implies a polynomial time algorithm to compute the Banzhaf indices. As a corollary, computing the Shapley value is \#P-complete for simple games represented by the set of minimal winning coalitions, Threshold Network Flow Games, Vertex Connectivity Games and Coalitional Skill Games.

\end{abstract}

\section{Introduction}

In this paper, we study the natural problem of computing the influence of edges in keeping an unweighted and undirected multigraph connected. 
Game theorists have studied notions of efficiency, fairness and stability extensively. 
Therefore, it is only natural that when applications in computer science and multiagent systems require fair and stable allocations, social choice theory and cooperative game theory provide appropriate foundations. 
For example, a network administrator with limited resources to maintain the links in the network may decide to commit resources to links according to their connecting ability. 
A spy network comprises communication channels.
In order to intercept messages on the channels, resources
may be utilized according to the ability of a channel to connect all groups. 
In a social network, we may be interested in checking which connections 
are more important in maintaining connectivity and hence contribute more to social welfare.

Our model is based on undirected, unweighted and connected
multigraphs.
All the nodes are treated equally, and the importance of a
edge is based solely on its ability to connect all the nodes. 
Using undirected edges is a reasonable assumption in many cases. 
For example, in a social network, relations are usually mutually formed. 


We use a multigraph as a succinct representation of a simple coalitional game 
called the \emph{spanning connectivity game (SCG)}. 
The players of the game are the edges of the multigraph.
The importance of an edge is measured by computing its voting
power index in the game. 
Voting power indices including the Banzhaf index and Shapley-Shubik
index are standard ways to compute the importance of a
player in a coalitional voting game.  Intuitively, the
Banzhaf value is the number of coalitions in which a player
plays a critical role and the Shapley-Shubik index is the
proportion of permutations for which a player is pivotal.



The whole paper is concerned with computing solutions for
SCGs.
In Section~\ref{SectionNetwork1RelWork}, a summary of related work is given. 
In Section~\ref{SectionNetwork1Prel}, preliminary
definitions related to graph theory and coalitional games
are given, and we define SCGs. 
Section~\ref{SectionNetwork1Indices} presents hardness results for computing 
Banzhaf values and Shapley-Shubik indices. 
In Section~\ref{SectionNetwork1Pos}, positive computational results 
for Banzhaf values and Shapley-Shubik indices are provided
for certain graph classes. 
Section~\ref{SectionNetwork1other} presents a polynomial-time algorithm 
to compute Holler indices and Deegan-Packel indices. 
In Section~\ref{SectionNetwork1Conc}, a summary of results is given and future work is discussed.

\section{Related work}\label{SectionNetwork1RelWork}

Power indices such as the Banzhaf and Shapley-Shubik indices have been extensively 
used to gauge the power of a player in different coalitional games such as 
weighted voting games~\cite{DBLP:conf/aaim/FaliszewskiH08} and corporate 
networks~\cite{DBLP:journals/eor/CramaL07}.  
These indices have recently been used in network flow games
\cite{DBLP:conf/atal/BachrachR07}, where the edges in the
graph have capacities and the power index of an edge
signifies the influence that an edge has in enabling a flow from the source to the sink. 
Voting power indices have also been examined in vertex
connectivity games \cite{DBLP:conf/atal/BachrachRP08} on
undirected, unweighted graphs; there the players are nodes,
which are partitioned into primary, standard, and backbone classes.

The study of cooperative games in combinatorial domains
is widespread in operations research~\cite{RePEc:spr:topjnl:v:9:y:2001:i:2:p:139-199,curielBool}.
Spanning network games have been examined previously
\cite{RePEc:spr:jogath:v:27:y:1998:i:4:p:467-500,RePEc:spr:jogath:v:21:y:1993:i:4:p:419-27}
but they are treated differently, with weighted graphs and \emph{nodes} as players (not
\emph{edges}, as here). 
The SCG is related to the all-terminal reliability model, a
non-game-theoretic model that is relevant in broadcasting~\cite{Valiant1979,4335422}. 
Whereas the reliability of a network concerns the overall
probability of a network being connected, 
this paper concentrates on resource allocation to the edges. 
A game-theoretic approach can provide fair and stable
outcomes in a strategic setting.


\section{Preliminaries}\label{SectionNetwork1Prel}


\subsection{Graph Theory}

\begin{definition}
A \emph{multigraph} $G:=(V,E,s)$ consists of a simple
underlying graph $(V,E)$ with a multiplicity function
$s:E\mapsto \mathbb{N}$ where $\mathbb{N}$ is the set of natural numbers excluding $0$. 
Let $|V|=n$ and $|E|=m$. 
For every underlying edge $i\in E$, we have $s_i$ edges in the multigraph. 
The multigraph has a total of $M=\sum_{i\in E}s_i$ edges.  
\end{definition}

\begin{definition}
A \emph{subgraph} $G'=(V',E')$ of a graph $G=(V,E)$ is a graph where $V'$ is a subset of $V$ and $E'$ is a subset of $E$ such that the vertex set of $E'$ is a subset of $V'$. 
A subgraph $H$ is a \emph{connected spanning subgraph} of a graph $G$ if it is connected and has the same vertex set as $G$. 
\end{definition}



\subsection{Coalitional Game Theory}

\begin{definition}
A \emph{simple voting game} is a pair $(N,v)$ with characteristic function $v:2^N \rightarrow \{0,1\}$ where $v(\emptyset)=0$, $v(N)=1$ and $v(S)\leq v(T)$ whenever $S \subseteq T$. A coalition $S \subseteq N$ is \emph{winning} if $v(S)=1$ and \emph{losing} if $v(S)=0$. A simple voting game can alternatively be defined as $(N,W)$ where $W$ is the set of winning coalitions. 
\end{definition}

For the sake of brevity, we will abuse the notation to sometimes refer to game $(N,v)$ as $v$.
For each connected multigraph $(V,E,s)$, we define the SCG, \emph{spanning
connectivity game}, $(E,v)$ with players $E$ and valuation
function $v$, defined as follows for $S\subseteq E$: 
\begin{displaymath}
v(S)= \left \{ \begin{array}{ll} 1, & \textrm{if there exists a spanning tree $T=(V,E')$ such that $E'\subseteq S$}\\
0, & \textrm{otherwise} \end{array} \right.
\end{displaymath}

It is easy to see that the SCG $(E,v)$ is a simple game because the outcome is binary, $v$ is monotone, $v(\emptyset)=0$ and $v(E)=1$. We consider power indices and cooperative game solutions for the edges in the SCG.

\begin{definition}
A player $i$ is \emph{critical} in a coalition $S$ when $v(S)=1$ and $v(S \setminus \{i\})=0$. For each $i \in N$, we denote the number of coalitions in which $i$ is critical in game $v$ by the \emph{Banzhaf value} ${{\eta}_{i}}(v)$. The \emph{Banzhaf Index} of player $i$ in game $v$ is 
$$\beta_i = \frac{{{\eta}_{i}}(v)}{{\sum}_{i \in
N}{{\eta}_{i}}(v)}\ .$$ 
\end{definition}

The \emph{Shapley-Shubik index} is the proportion of permutations for which a player is \emph{pivotal}. For a permutation $\pi$ of $N$, the $\pi(i)$th player is pivotal if coalition $\{\pi(1),\ldots, \pi(i-1)\}$ is losing but coalition $\{\pi(1),\ldots, \pi(i)\}$ is winning.

\begin{definition}
The \emph{Shapley-Shubik (SS) value} is the function $\kappa$ that assigns to any simple game $(N,v)$ and any voter $i$ a value $\kappa_{i}
(v)$ where $$\kappa_{i} = \sum_{S \subseteq N} (|S|-1)!(n-|S|)!(v(S)- v(S\setminus \{i\})).$$ The \emph{Shapley-Shubik (SS) index} of $i$ is defined by $$\phi_i=\frac{{\kappa}_{i}}{n!}.$$
\end{definition}

The Banzhaf index and the Shapley-Shubik index are the normalized versions of the Banzhaf value and the Shapley-Shubik value respectively. Since the denominator of the Shapley-Shubik index is fixed, computing the Shapley-Shubik index and Shapley-Shubik value have the same complexity. This is not necessarily true for the Banzhaf index and Banzhaf value.

\section{Complexity of computing power indices}\label{SectionNetwork1Indices}


We define the problems of computing the power indices of the edges in the SCG.
For any power index X (e.g. Banzhaf value, Banzhaf index, Shapley-Shubik index etc.) we define the problem SCG-X as follows:\\

\noindent
\textbf{Problem}: SCG-X\\
\textbf{Instance}: Multigraph $G$\\
\textbf{Output}: For the SCG corresponding to $G$, compute X for all the edges\\



We represent a communication network as a
multigraph, where an edge represents a connection that may or may
not work.
An edge is said to be {\em operational} if it works.  
For a given graph $G$, the reliability
$Rel(G,\{p_i\})$ of $G$ is the probability that 
the operational edges form a connected spanning subgraph, 
given that each edge is operational with probability $p_i$
for $i=1,\ldots m$.\\

\noindent
\textbf{Problem}: Rational Reliability Problem\\
\textbf{Instance}: Multigraph $G$ and $p_i\in \mathbb{Q}$ for all $i$, $1\leq i\leq m$\\
\textbf{Output}: Compute $Rel(G,\{p_i\})$\\

A special case of the 
reliability problem is when 
every edge has the same probability $p$ of being
operational. This is called the \emph{Functional Reliability Problem}. A connected spanning subgraph with $i$
edges will occur with probability $p^i{(1-p)}^{m-i}$. 

\begin{definition}\label{RelPolyDef}
Let $N_i$ be the number of connected spanning subgraphs with $i$ edges. Then the required output of the Functional Reliability Problem is the \emph{reliability polynomial} $$Rel(G,p)=\sum_{i=0}^{m}{N_i}p^i{{(1-p)}^{m-i}}.$$
\end{definition}
\noindent
\textbf{Problem}: Functional Reliability Problem\\
\textbf{Instance}: Multigraph $G$\\ 
\textbf{Output}: Compute the coefficients $N_i$ of the reliability polynomial for all $i$, $1\leq i\leq m$.\\

Ball~\cite{4335422} points out that an algorithm to solve the Rational Reliability Problem can be used as a sub-routine to compute all the coefficients for the Functional Reliability Problem. Moreover he proved that computing the general coefficient $N_i$ is NP-hard and therefore computing the rational reliability of a graph is NP-hard.
As we will see in Section~\ref{SectionNetwork1Pos}, reliability problems have connections with computing power indices of SCG.
We first prove that SCG-BANZHAF-VALUE is \#P-complete.

\begin{proposition}
SCG-BANZHAF-VALUE is \#P-complete even for simple, bipartite and planar graphs.
\end{proposition}
\begin{proof}
We present a reduction from the problem of counting connected spanning subgraphs.
SCG-BANZHAF-VALUE is clearly in \#P because a connected
spanning subgraph can be verified in polynomial time. 
It is known that counting the total number of connected spanning subgraphs is 
\#P-complete even for simple, bipartite and planar graphs(~\cite{SurveysComb1997}, p. 305). 
We now reduce the problem of computing the total number of
connected spanning subgraphs to solving SCG-BANZHAF-VALUE. Take $G=(V,E)$ with $n$ nodes and $m$ edges. 
Transform graph $G$ into $G'=(V\cup\{n+1\},E\cup\{m+1\})$ by taking any node and connecting it to a new node via a new edge. 
Then the number of spanning subgraphs in $G$ is equal to the Banzhaf value of edge $m+1$ in graph $G'$. 
This shows that SCG-BANZHAF-VALUE is \#P-complete.  
\end{proof}

Similarly, SCG-SS is \#P-complete.

\begin{proposition}\label{PropSCG-SS}
SCG-SS is \#P-complete even for simple graphs.
\end{proposition}
\begin{proof}

Let $N_i$ be the number of connected spanning subgraphs of $G$ with $i$ edges. We know that computing $N_i$ is NP-hard~\cite{4335422}. 
We show that if there is an algorithm polynomial 
in the number of edges to compute the Shapley-Shubik index 
of all edges in the graph, then each $N_i$ can be computed 
in polynomial time. 

We obtain graph $G_0$ by the following transformation: for some node $v \in V(G)$, we link it by a new edge $x$ to a new node $v_x$. Then, by the definition of the Shapley-Shubik value,  
$\sum_{r=0}^{m}r!N_r(|E(G)|-r)!=\sum_{r=0}^{m}r!N'_r= \kappa_{x}(G_0)$, where 
we write $N'_r$ for $N_r(m-r)!$, for all $r$.

Similarly we can construct $G_i$ by adding a path $P_i$ of length $i$ to $v_x$ where $P_i$ has no edge or vertex intersection with $G$.
Therefore 
\begin{equation}\label{ShapleyEquation}
\sum_{r=0}^{m}(r+i)!N'_r=\kappa_{x}(G_i).
\end{equation}

For $i=0,\ldots, m$, we get an equation of the form of \eqref{ShapleyEquation} for each $G_i$. The left-hand side of the set of equations can be represented by an $(m+1)\times (m+1)$ matrix $A$ where $A_{ij}={(i+j-2)!}$. The set of equations is independent because $A$ has a non-zero determinant of ${(1!2!\cdots m!)}^2$ (see e.g. Theorem 1.1~\cite{BacherPascal}).
If there is a polynomial time algorithm to compute the Shapley-Shubik index of each edge in a simple graph, then we can compute the right-hand side of each equation corresponding to $G_i$.

The biggest possible number in the
equation is less than $(2m)!$ and can be represented
efficiently. According to Stirling's formula, $m!\approx
\sqrt{2\pi m}\left(\frac{m}{e}\right)^m$, the number $(2m)!$
can be represented by $km(log~m)$ bits where $k$ is a constant.
We can use Gaussian elimination to solve the set of linear
equations in $\mathcal{O}(m^3)$ time. Moreover, each number
that occurs in the algorithm can also be stored in a number
of bits quadratic of the input size (Theorem
4.10~\cite{korte06}). 
Therefore SCG-SS is \#P-complete.
  
\end{proof}

A representation of a simple game is considered \emph{reasonable} if, for a simple game $(N,v)$, the new game $(N\cup\{x\},v')$ where $v(S)=1$ if and only if $v'(S\cup\{x\})=1$, can also be represented. Then the proof technique in Proposition~\ref{PropSCG-SS} can be used to show that for any reasonable representation of the simple game, a polynomial time algorithm to compute the Shapley-Shubik indices implies a polynomial time algorithm to compute the Banzhaf indices. 
This answers (positively) the question from \cite{Azizcomsoc2008} of whether computing Shapley-Shubik indices for a simple game represented by the set of minimal winning coalitions is NP-hard. 
As a corollary, we also strengthen or settle the complexity of a number of coalitional games. The proof in Proposition~\ref{PropSCG-SS} can be slightly modified to prove that computing the Shapley-Shubik index (Shapley value in case of non-simple games) is $\#$P-complete for a number of games:


\begin{proposition}
Computing Shapley value is $\#$P-complete for 
\begin{enumerate}
\item Simple game represented by its minimal winning coalitions
\item Threshold Network Flow Games~\cite{Bachracha:2008:JAAMAS}
\item Vertex Connectivity Games~\cite{DBLP:conf/atal/BachrachRP08}
\item STSG (Single Task Skill Game), TCSG (Task Count Skill Game), WTSG (Weighted Task Skill Game), TCSG-T (Task Count Skill Game with thresholds) and WTSG-T (Weighted Task Skill Game with thresholds)~\cite{DBLP:conf/atal/BachrachR08}
\end{enumerate}
\end{proposition}
\begin{proof}
For the given games, computing Banzhaf values is $\#$P-complete. It is easy to see that the games Threshold Network Flow Games, Vertex Connectivity Games, STSG (Single Task Skill Game), TCSG-T (Task Count Skill Game with thresholds) and WTSG-T (Weighted Task Skill Game with thresholds) are simple games with reasonable representations. Also, TCSG (Task Count Skill Game) and WTSG (Weighted Task Skill Game) are generalizations of the STSG (Single Task Skill Game).
\end{proof}

\section{Polynomial time cases}\label{SectionNetwork1Pos}

In this section, we present polynomial time algorithms to compute voting power indices for restricted graph classes including graphs with bounded 
treewidth. We first consider the trivial case of a tree.
If the graph $G=(N,E)$ is a tree then there is a total of
$n-1$ edges and only the grand coalition of edges is a winning coalition. Therefore a tree is equivalent to a unanimity game. This means that each edge has a Banzhaf index and Shapley-Shubik index of $\frac{1}{n-1}$. In the case of the same tree structure but with multiple parallel edges, we refer to this multigraph as a \emph{pseudo-tree}.


\begin{proposition}
Let $G=(N,E,s)$ be a pseudo-tree such that the underlying edges are $1,\ldots, m$ with multiplicities $s_1,\ldots, s_{m}$. Then,

\begin{equation}\label{pseudoEqn}
\eta_{i_1}={\prod_{\substack{j=1\\j\neq i}}^{m}} (2^{s_j}-1), {\rm\ and\ so\ } \beta_{i_1}=\frac{\eta_{i_1}}{\sum_{k=1}^{m}s_k \eta_{k_1}}.
\end{equation}
\end{proposition}
\begin{proof}
Note that $m=n-1$ in this case.
Suppose edge $i_1$ is a parallel edge corresponding to edge $i$ in the underlying graph.
Edge ${i_1}$ is critical for a coalition $C$ if the coalition $C$ contains no edges parallel to $i_1$ but contains at least one sub-edge corresponding to each edge other than $i$. The number of such coalitions is $\prod_{\substack{j=1\\j\neq i}}^{m}(2^{s_j}-1)$, which gives~\eqref{pseudoEqn}.

\end{proof}

\begin{proposition}\label{PropSStree}
Let $G=(N,E,s)$ be a pseudo-tree such that the underlying edges are $1,\ldots, m$ with multiplicities $s_1,\ldots, s_m$ where $s=\sum_{i=1}^m s_i$. Then the Shapley-Shubik indices can be computed in time polynomial in the total number of edges.
\end{proposition}
\begin{proof}
Denote by $e_r$ the coefficient of $x^r$ in 

$$\prod_{\substack{1\leq j\leq n-1\\{j\neq i}}}({{{(1+x)}^{s_j}}-1}).$$

Then $e_r$ is the number of coalitions with $r$ edges which include at least one parallel edge for each underlying edge $j$ except $i$. 
Then, by definition of the Shapley-Shubik value, for $1\leq k\leq s_i$,

$$\kappa_{i_k}(G)=\sum_{r=n-2}^{s-s_i}e_r r!(s-r-s_i)!.$$

Thus, the Shapley-Shubik indices can be computed in time polynomial in the total number of edges.

\end{proof}

We now consider graphs with bounded treewidth. Note that trees and pseudo-trees hve treewidth 1.

\begin{definition}
For a graph $G = (V,E)$, a \emph{tree decomposition} is a pair $(X,T)$, where $X = \{X_1, ..., X_n\}\subset 2^V$, and $T$ is a tree whose nodes are the subsets $X_i$ with the following properties:
\begin{enumerate}
\item $\bigcup_{1\leq i\leq n}X_i=V$
\item For every edge $(v,w)\in E$, there is a subset $X_i$ that contains both $v$ and $w$.
\item If $X_i$ and $X_j$ both contain a vertex $v$, then all nodes $X_z$ of the tree in the path between $X_i$ and $X_j$ also contain $v$.
\end{enumerate}
The \emph{width} of a tree decomposition is the size of its largest set $X_i$ minus one. The \emph{treewidth} $tw(G)$ of a graph $G$ is the minimum width among all possible tree decompositions of $G$.
\end{definition}

\begin{proposition}\label{PolynomialBanzhafIndexInGraph}
If the reliability polynomial defined in Definition~\ref{RelPolyDef} can be computed in polynomial time, then the following problems can be computed in time polynomial in the number of edges:
\begin{enumerate}
\item the number of connected spanning subgraphs; 
\item the Banzhaf indices of edges.
\end{enumerate}
\end{proposition}
\begin{proof}
We deal with each case separately.
\begin{enumerate}
\item  By definition, $N_i$ is the number of connected spanning subgraph with $i$ edges. If all coefficients $N_i$ are computable in polynomial time, then the total number of connected spanning subgraphs $\sum_{i=0}^m N_i$ is computable in polynomial time. 
\item We know that $\eta_i(G)=2\omega_i(G)-\omega(G)$~(See~\cite{BanzhafMath1979}) where $\omega(G)$ is equal to the total number of winning coalitions and 
$\omega_i(G)$ is the number of winning coalitions including
player $i$. Consider the graph $G$ where the probability of
edge $i$ being operational is set to $1$ whereas the
probability of other edges being operational is set to
$0.5$. Then the reliability of the graph being connected is
equal to the ratio of the number of connected spanning
subgraphs that include edge $i$ to $2^{M-1}$, the total number of
subgraphs that include~$i$. Therefore, $\omega_i(v)$ the number of
connected spanning subgraphs including edge $i$ can be computed in polynomial time too.
\end{enumerate}
\end{proof}

\begin{corollary}
Banzhaf indices of edges can be computed in polynomial time for graphs with bounded treewidth.
\end{corollary}
\begin{proof}
This follows from the polynomial time algorithm to compute the reliability of a graph with treewidth $k$ for some fixed $k$~\cite{65065}. 
\end{proof}

\begin{definition}
Let $G=(V,E)$ be a graph with source $s$ and sink $t$. Then $G$ is a \emph{series-parallel} graph if it may be reduced to $K_2$ by a sequence of the following operations:
\begin{enumerate}
\item replacement of a pair of parallel edges by a single edge that connects their common endpoints;
\item replacement of a pair of edges incident to a vertex of degree 2 other than $s$ or $t$ by a single edge.
\end{enumerate}
\end{definition}

Graphs with bounded treewidth can be recognized in polynomial time~\cite{37183}.  
Series-parallel graphs and $2$-trees are well-known classes of graphs with constant treewidth. 
Other graph classes with bounded treewidth are cactus graphs and outer-planar graphs.
We see that whereas computing Banzhaf values of edges in general SCGs is NP-hard, important graph classes can be recognized and their Banzhaf values computed in polynomial time. 

When edges have special properties, their power indices may be easier to compute. 
We define a \emph{bridge} in the graph to be an edge whose removal results in the graph being disconnected. 
A graph class is \emph{hereditary} if for every graph in the class, every subgraph is also in the class. 

\begin{proposition}
If graph $G$ belongs to a hereditary graph class, for which the reliability polynomial of a graph can be computed in polynomial time, then the Shapley-Shubik index of a bridge can be computed in time polynomial in the total number of edges.
\end{proposition}
\begin{proof}
Let graph $G=(V,E)$ be a graph where edge $k$ is a bridge
which connects two components $A=(V_A,E_A)$ and
$B=(V_B,E_B)$. Then $|E|=|E_A|+|E_B|+1$. If the 
reliability polynomial of $G$ can be computed in polynomial
time, then the 
reliability polynomial for each of the components $A$ and $B$ can be computed. Then the Shapley-Shubik index of player $k$ is: 
$$\phi_k(G)= \frac{\sum_{i=|V_A|-1}^{|E_A|} \sum_{j=|V_B|-1}^{|E_B|}N_i(A)N_j(B)(i+j)!(|E_A|+|E_B|-i-j)!}{|E|!}.
$$ 
\end{proof}

\begin{figure}
  \begin{center}
      \includegraphics[scale=0.2]{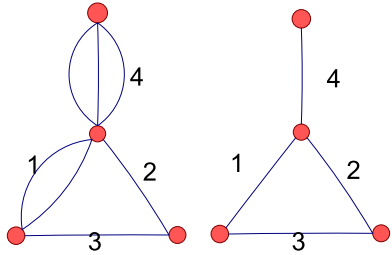}
      
  \end{center}
  \caption{Multigraph and its underlying graph\label{GraphExampleBoth}\normalsize}
  \end{figure}

Our next result is that if the reliability of a simple graph can be computed then the Banzhaf indices of the corresponding multigraph can be computed. 
A naive approach would be to compute the Banzhaf values of each edge in a simple graph and then, for the corresponding parallel edges in the multigraph, divide the Banzhaf value of the overall edge by the number of parallel edges. 
However, as the following example shows, this approach is incorrect:

\begin{example}\label{ProblemOfSimpleToMulti}
Let $G=(V,E,s)$ be the multigraph in Figure~\ref{GraphExampleBoth}. Then, $\eta_{4_1}(v_G)=10$, $\eta_{1_1}(v_G)=14$, and $\eta_{2}(v_G)=\eta_{3}(v_G)=28$. Therefore $\beta_{4_1}(v_G)=\frac{10}{3\times 10 + 2\times 14 + 28+28}=\frac{5}{57}$. Moreover, $\beta_{1_1}(v_G)=\frac{7}{57}$ and $\beta_{2}(v_G)=\beta_{3}(v_G)=\frac{14}{57}$. If we examine the underlying graph of $G'$ in Figure~\ref{GraphExampleBoth}, then $\eta_{4}(v_G')=4$ and $\eta_{1}(v_G')=\eta_{2}(v_G')=\eta_{3}(v_G')=2$ 
giving $\beta_4(G')=2/5$ and $\beta_i(G')=1/5$ for $i=1,2,3$.
Therefore, the Banzhaf values of edges in the underlying graph do not give a direct way of computing the Banzhaf values in the multigraph. 
\end{example}

\begin{lemma}\label{MultigraphRelLEMMA}
If there is an algorithm to compute the reliability of the underlying simple graph, then the algorithm can be used to compute the reliability of the corresponding multigraph. 
\end{lemma}
\begin{proof}
Let $G=(V,E,s)$ be a multigraph in which there are $s_i$ parallel edges $i_1, \ldots, i_{s_i}$ corresponding to edge $i$. Let $p_{{i_j}}$ be the probability that the $j$th parallel edge of edge $i$ is operational. In that case $Rel(G,p)$ is equal to $Rel(G',p')$, where $G'$ is the corresponding simple graph of $G$ and the probability $p_{i}$ that edge $i$ is operational is $1-\prod_{j=1}^{s_i}(1-p_{i_j})$. 
\end{proof}

We now prove in Proposition~\ref{reliabilityBanzhaf} that if there is an algorithm to compute the reliability of the underlying simple graph $G$, then it can be used to compute the Banzhaf indices of the edges in the corresponding multigraph of $G$. 
It would appear that the proposition follows directly from Lemma~\ref{MultigraphRelLEMMA} and Proposition~\ref{PolynomialBanzhafIndexInGraph}. However, one needs to be careful that the reliability computed is the reliability of the overall graph. 
Example~1 
 shows that computing the Banzhaf values of the edges in the underlying simple graph does not directly provide the Banzhaf values of the parallel edges in the corresponding graph. 

\begin{proposition}\label{reliabilityBanzhaf}
For a multigraph $G$ and edge $i$, let $G'$ be the multigraph where all the other edges parallel to edge $i$ are deleted. Then if the reliability of $G'$ can be computed in polynomial time, then the Banzhaf value of edge $i$ in $G$ can be computed directly by analysing $G'$.

\end{proposition}
\begin{proof}
Recall that $G$ is a multigraph with a total of $M$ edges.
Given an algorithm to compute the reliability of $G'$, we provide an algorithm to compute the Banzhaf values of the parallel edges of edge $i$ in $G$. For graph $G'$, set the operational probabilities of all edges to $0.5$ except $i$ which has an operation probability of $1-{0.5}^{s_i}$.
and compute the overall reliability $r(G')$ of the graph. 
Then, by Lemma~\ref{MultigraphRelLEMMA}, $\omega(G)$ is $2^M{r(G')}$.

Now for $G'$, set the operational probabilities of all edges to $0.5$ except $i$ which has an operation probability of $1$. 
Let the reliability of $G'$ with the new probabilities be $r'(G')$. 
We see that $r'(G')$ is equal to $\omega_i(G')/{2^{M-s_i}}$.
Then $\omega_i(G) = 2^{s_i-1}\omega_i(G') = 2^{M-1}r'(G')$.
The Banzhaf value of $i$ is then $2\omega_i(G) - \omega(G)$.
A similar approach gives Banzhaf values of other edges from which all the Banzhaf indices can be computed.

\end{proof}




\section{Other power indices}\label{SectionNetwork1other}

Apart from the Banzhaf and Shapley-Shubik indices, there are other indices which are also used. Both the Deegan-Packel index~\cite{deeganpackel1978} and the Holler index~\cite{Holler1982} are based on the notion of minimal winning coalitions. Minimal winning coalitions are significant with respect to coalition formation. The Holler index, $H_i$ of a player $i$ in a simple game is similar to the Banzhaf index except that only swings in minimal winning coalitions contribute toward the Holler index. 

\begin{definitions}
Let $M_i$ be $\{S\in W^m: i \in S\}$.
We define the \emph{Holler value} as $|M_i|$. 
The \emph{Holler index} 
(also called the \emph{public good index}) is defined by 
$$H_i(v)=\frac{|M_i|}{\sum_{j\in N}|M_j|}.$$ 
The \emph{Deegan Packel index} for player $i$ in voting game $v$ is defined by 
$$D_i(v)={\frac{1}{|W^m|}}{\sum_{S\in M_i}{\frac{1}{|S|}}}.$$
\end{definitions}

\begin{proposition}
For SCGs corresponding to multigraphs, Holler indices and Deegan-Packel indices can be computed in polynomial time.
\end{proposition}
\begin{proof}
We use the fact that the number of trees in a multigraph can be computed in polynomial time, which follows from \emph{Kirchhoff's matrix tree theorem}~\cite{citeulike:366370}. Given a connected graph $G$ with $n$ vertices, let $\lambda_1,\lambda_2,...,\lambda_{n - 1}$ be the non-zero eigenvalues of the Laplacian matrix of $G$ (the Laplacian matrix is the difference of the degree matrix and the adjacency matrix of the graph). Kirchhoff proved that the number of spanning trees of $G$ is equal to any cofactor of the Laplacian matrix of $G$~\cite{citeulike:366370}: $t(G)=\frac{1}{n}\lambda_1\lambda_2\cdots\lambda_{n-1}$. 
So now that we have a polynomial-time method to compute the
number of spanning trees $t(G)$ of graph $G$, we claim this
is sufficient to compute the Holler values of the edges. If
an edge $i$ is a bridge, then it is present in every
spanning tree and its Holler value is simply the total
number of spanning trees.  
If $i$ is not a bridge then $M_i=t(G)- t(G\setminus e)$. 
Moreover, since the size of every minimal winning coalition 
is the same, namely $(n-1)$, the Holler indices and Deegan Packel Indices 
coincide for an SCG.
\end{proof}

\section{Conclusion}\label{SectionNetwork1Conc}

This paper examined fairness-based cooperative game
solutions of SCGs, for allocating resources to edges. 
In another recent paper, we have also looked at the computation of stability based cooperative game solutions of SCGs. A polynomial time algorithm is presented to compute the nucleolus. This is a surprising result considering that the standard power indices are NP-hard to compute and also that the SCG is not convex in general. Therefore, the nucleolus may be a better alternative for resource allocation in SCGs.

We looked at the exact computation of power indices. 
In \cite{1402354}, an optimal randomized algorithm to compute Banzhaf indices and Shapley-Shubik indices with the required confidence interval and accuracy is presented. Since the analysis in \cite{1402354} is not limited to a specific representation of a coalitional game, it can be used to approximate Banzhaf indices and Shapley-Shubik indices in SCGs.

The results of the paper are summarized in Table~\ref{tableSpanning}. 
This framework can be extended to give an ordering on 
the importance of nodes in the graph~\cite{networkalgorithms}. 
To convert a resource allocation to edges to one on nodes, 
the payoff for an edge is divided equally between its two
adjacent nodes. 
The total payoff of a node is the sum of the payoffs it
gets from all its adjacent edges. 
This gives a way to quantify and compare the centrality 
or connecting role of each node. 
It will be interesting to understand the properties 
of such orderings, especially for unique cooperative solution 
concepts such as the nucleolus, Shapley-Shubik and Banzhaf indices.

The complexity of computing the Shapley-Shubik index for an SCG 
with a graph of bounded treewidth is open. 
If this problem is NP-hard, it will answer the question posed 
in the conclusion of \cite{1402354}
 on whether there are any domains where computing one of 
the Banzhaf index and Shapley-Shubik index is easy, whereas 
computing the other is hard.

%
\begin{table*}
\centering
\caption{Complexity of SCGs}
\begin{tabular}{|l|c|c|} \hline
Problem&Input&Complexity\\ \hline
\hline
SCG-BANZHAF-VALUE&Simple, bipartite, planar graph&\#P-complete\\ \hline
SCG-BANZHAF-INDEX&Simple graph&?\\ \hline
SCG-BANZHAF-(VALUE/INDEX)&Multigraph with bounded treewidth&P\\ \hline
SCG-SS&Multigraph&\#P-complete\\ \hline
SCG-SS&Multigraph with bounded treewidth&?\\ \hline
SCG-H-(VALUE/INDEX)&Multigraph&P\\ \hline
SCG-DP-(VALUE/INDEX)&Multigraph&P\\ \hline
\end{tabular}
\label{tableSpanning}
\end{table*}
\section{Acknowledgements}

Partial support for this research was provided by DIMAP 
(the Centre for Discrete Mathematics and its Applications), 
which is funded by the UK EPSRC under grant EP/D063191/1. 
Rahul Savani also received partial support from EPSRC grant
EP/D067170/1.
Haris Aziz would also like to thank the Pakistan National 
ICT R\&D Fund for funding his research.

\bibliographystyle{plain}

{\bf HARIS AZIZ}
\\
{\it Department of Computer Science, University of Warwick, 
Coventry CV4 7AL, United Kingdom.\\
haris.aziz@warwick.ac.uk.}

{\bf ODED LACHISH}
\\
{\it Department of Computer Science, University of Warwick, 
Coventry CV4 7AL, United Kingdom.\\
oded@dcs.warwick.ac.uk.}

{\bf MIKE PATERSON}
\\
{\it Department of Computer Science, University of Warwick, 
Coventry CV4 7AL, United Kingdom.\\
msp@dcs.warwick.ac.uk.}

{\bf RAHUL SAVANI}
\\
{\it Department of Computer Science, University of Warwick, 
Coventry CV4 7AL, United Kingdom.\\
rahul@dcs.warwick.ac.uk}

\end{document}